\documentclass[10pt,letterpaper]{article}
\usepackage{opex3}

\usepackage{amsmath}    
\usepackage{graphicx} 
\usepackage{verbatim}   
\usepackage{color}      
\usepackage{hyperref}   
\usepackage{amssymb}
\usepackage{dsfont}

\newcommand{\bra}[1]{\left\langle#1\right\vert}
\newcommand{\ket}[1]{\left\vert#1\right\rangle}

\begin{document}

\title{Tomography of Spatial Mode Detectors}
\author{I.B.Bobrov, E.V.Kovlakov, A.A.Markov, S.S.Straupe, S.P.Kulik}

\address{Faculty of Physics, M.V.Lomonosov Moscow Sate University}
\email{straups@yandex.ru}

\date{\today}

\begin{abstract}
Transformation and detection of photons in higher-order spatial modes usually requires complicated holographic techniques. Detectors based on spatial holograms suffer from non-idealities and should be carefully calibrated.
We report a novel method for analyzing the quality of projective measurements in spatial mode basis inspired by quantum detector tomography. It allows us to calibrate the detector response using only gaussian beams. We experimentally investigate the inherent inaccuracy of the existing methods of mode transformation and provide a full statistical reconstruction of the POVM (positive operator valued measure) elements for holographic spatial mode detectors.
\end{abstract}



\section{Introduction}
Analyzing mode content of a spatially multimode beam is an important primitive in classical and quantum optics, where spatial modes are gaining more and more attention as a convenient degree of freedom for information encoding and multiplexing tasks. For example use of mode division multiplexing allows to significantly increase the information transmission rate in free space \cite{wang2012terabit} and fiber optical \cite{bozinovic2013terabit} communications. In quantum optics spatial degrees of freedom of single photons and correlated photon pairs, most prominently the orbital angular momentum of helical Laguerre-Gaussian modes \cite{molina2007twisted}, are used in multidimensional quantum communication protocols \cite{vaziri2002qutrits, molina2001management, langford2004qutrits, groblacher2006ecryptography, mafu2013MUB} and high-dimensional entanglement experiments \cite{mair2001entanglement, vaziri2003concentration, walborn2004entanglement, oemrawsingh2005fractional, jack2010entanglement, dada2011experimental, fickler2012entanglement}. Other mode functions, such as Hermite-Gaussian modes may find similar applications as well \cite{walborn2005HG, walborn2012generalizedHG, straupe2011Schmidt, miatto2012cartesian}.

While generation of higher-order Laguerre-Gaussian \cite{arlt1998production, matsumoto2008generation, ando2009mode, kennedy2002diffractive}, Hermite-Gaussian and even quite exotic modal superpositions \cite{bentley2006Ince--Gaussian, maurer2007tailoring, siviloglou2007observation, zhang2012Mathieu} is a well established technique, constructing an efficient higher-order spatial mode filter is not an easy task. One of the most attractive options is to reverse the generation process and use specially designed phase holograms to transform the higher-order beam into a Gaussian one, which can be easily filtered with a single mode fiber. In quantum optical language such filer, if it was perfect, will correspond to an ideal projective measurement. Obviously, any real mode transforming hologram will not produce a perfect Gaussian beam \cite{qassim2014limitations}, so the measurement will not be described by an exact projector, but rather by a more complicated POVM. If one wants to correctly interpret the measurement results, a method for reconstructing the real POVM elements of holographic spatial mode filters is desired. Here we present such a method and realize it experimentally for several types of commonly used holograms. These results provide a quantitative measure of inherent detector non-ideality, which should be taken into account in all experiments involving spatial mode-sensitive detection, be it an experiment on quantum state tomography, classical mode division multiplexing or anything else.

\section{Holographic spatial detectors}

Let us first consider consider beam transformations by phase-only holograms. In a typical experimental situation an initially Gaussian beam  with complex field amplitude distribution in the plane of the hologram described by $\mathcal{E}_{in}(x,y)=\mathcal{E}_0\exp(-(x^2+y^2)/2w^2)$, is incident on a planar phase-only spatial light modulator (SLM). We will assume the polarization to be fixed and consider a scalar problem.

The hologram consists of some smooth phase profile superposed with a blazed grating of variable modulation depth. Since the amplitude in the first diffraction order of the grating depends on the modulation depth, such a hologram acts as both phase and amplitude modulator \cite{Kirk1977Filter, Davis99Amplitude}. The phase profile corresponding to such hologram in its most general form is
\begin{equation}\varphi (x, y)=M(x, y)\,\mathrm{mod}_{2\pi}\!\left(F(x, y)+\frac{2\pi x}{\Lambda}\right),\end{equation}
where $x$ and $y$ are the coordinates in the hologram plane, $M(x, y)$ is the normalized grating modulation depth ($0\leq M \leq 1$), $F(x, y)$ is some function of phase and amplitude of the desired output field and $\Lambda$ is the diffraction blaze-grating period. The hologram should be designed to produce the desired output field distribution $\mathcal{E}_{out}(x,y) = A(x,y)\exp(i\Phi(x,y))$ in the first diffraction order of the grating. It was shown recently \cite{bolduc2013exact}, that the \emph{exact solution} to this problem is given by the following expressions for $M(x, y)$ and $F(x,y)$:
\begin{eqnarray}
\label{eq:ExactMask}
M(x, y) &=& \mathrm{sinc}(\pi(A(x, y)-1)), \nonumber \\
F(x, y) &=& \Phi(x, y) - \pi A(x, y).
\end{eqnarray}

If the output intensity corresponds to some eigenfunction of the paraxial wave equation. i.e. to a \emph{transverse mode} the described scheme acts as a programmable mode converter: it takes a fundamental gaussian beam as an input and outputs a higher-order mode, corresponding to the specific hologram, displayed on the SLM. Obviously, if amplitude modulation is used such mode transformer cannot be lossless even in principle -- some intensity has to go to the zeroth diffraction order. For the same reasons it cannot be completely inverted. However, numerous works (see for example \cite{mair2001entanglement, Syouji2010converter, jack2010entanglement, salakhutdinov2012full-field}) use the inversed mode transformer as a spatial mode detector/filter -- a device, that takes some superposition (or statistical mixture) of beams in different spatial modes as an input and outputs a gaussian beam with intensity proportional to the weight of a particular mode in the superposition (mixture). Indeed, since orthogonal modes correspond to orthogonal holograms, only the desired mode will be mapped to a gaussian output, which can then be filtered out with a single mode fiber. Of course, the fiber mode should be carefully matched to the output mode of the hologram. To minimize the detection losses one may consider modulating only the phase of the field, thus maximizing the diffraction efficiency in the first order, but sacrificing the output mode quality (see Fig.~\ref{fig:Efficiency} for comparison of efficiencies for two methods).

The described type of detectors is widely used in quantum optical experiments, where they realize (approximate) projective measurements in the basis of spatial modes. A natural question arises: how well does a real hologram followed by a single mode fiber approximates an ideal projective measurement? The quantitative answer to this question can be given by the procedure called \emph{quantum detector tomography} \cite{lundeen2008tomography}, which allows one to reconstruct the detector response to the input states from the chosen basis -- in our case, to the particular set of spatial modes.

The most commonly used transverse modes are Hermite-Gaussian (HG) and Laguerre-Gaussian (LG) modes. They are solutions of the paraxial wave equation in free space in cartesian and polar bases respectively. In what follows we will use HG modes as our preferable basis for the reasons described below. The (normalized) field distribution for a HG beam for infinite wavefront radius of curvature (at the beam waist) has the following form:
\begin{eqnarray}
\varphi_{mn}(x,y)&=&\sqrt{\frac{2}{\pi w^2 2^{m+n}n! m!}}\, \mathrm{H}_m\!\left(\frac{\sqrt{2} x}{w} \right) \mathrm{H}_n\!\left(\frac{\sqrt{2} y}{w} \right) \nonumber \\
&\times& \exp \left( - \frac{x^2+y^2}{w^2} \right),
\label{eq:HG}
\end{eqnarray}
here $\mathrm{H}_{m}(x)$ are Hermite polynomials, $x$ and $y$ -- the transverse coordinates and $w$ -- the beam waist.

\section{Detector tomography}

Although the spatial mode filters are not limited to experiments at the single photon level, it will be convenient for our purposes to use quantum-mechanical notation and describe the filter with a corresponding positive operator-valued measure (POVM). For example, let us consider an ideal mode filter, which can be tuned to project on any mode out of the set $\ket{\psi_n}$. The corresponding POVM will consist of one dimensional orthogonal projectors
\begin{equation}
\pi_n=\left|\psi_n \right\rangle \left\langle \psi_n \right|.
\end{equation}
The spatial state of the input field can be described by a density matrix $\rho$. Then the probability of detecting a photon after the filter projecting on the $n-$th mode (in the classical case -- partial intensity after the filter) will be given by the Born's rule:
\begin{equation}
P_{\rho, n}=\mathrm{Tr}\left(\rho \pi_n \right).
\end{equation}
If the input field is in a pure spatial mode $\ket{\psi_m}$, the probability distribution of the outputs reduces to $P_{m, n}=\delta_{n,m}$.

However, as argued above the real-world mode-filters are never ideal, so corresponding POVM will have a more complicated structure:
\begin{equation}
\label{eq:POVM}
\tilde{\pi}_n=\sum_{k,p}{\theta^{(n)}_{k,p}\ket{\psi_k} \bra{\psi_p}}.
\end{equation}
The coefficients $\theta^{(n)}_{k,p}$ are to be determined experimentally with an appropriate calibration procedure. Direct measurement of the coefficients is not a good option, since in that case one needs to generate beams in the spatial modes $\ket{\psi_m}$ from the chosen set. It means that the same technique will be used for both generation and measurement, making it impossible to distinguish the intrinsic measurement unideality from possible preparation errors. The solution is to use a well-defined and easy to prepare set of \emph{calibration states} and perform statistical reconstruction of POVM elements from the measured data \cite{feito2009measuring}.

Let us describe the detector in a basis of Hermite-Gaussian modes $\varphi_{mn}$. In this case a convenient choice of calibration states is given by displaced Gaussian beams with (normalized) amplitude given by:
\begin{equation}
\label{eq:DispalcedGaussian}
\varphi_{00}(x-d_i,y)=\sqrt{\frac{2}{\pi w^2}}\exp\left(-\frac{(x-d_i)^2+y^2}{w^2} \right),
\end{equation}
here index $i$ numerates the discrete set of shifts $d_i$ ($i=\{0,\ldots,D-1\}$). POVM description in terms of HG modes is convenient, since the two-dimensional problem reduces to two independent one-dimensional ones, corresponding to shifts in the horizontal and vertical directions. In the following we will consider a one-dimensional case. Generalization to a full two-dimensional POVM reconstruction is straightforward. Decomposing the displaced Gaussian function in a basis of HG modes one can find an analytical expression for the probabilities of ideal projective measurements:
\begin{eqnarray}
&&P_{d_i,n} = \langle d_i|\pi_n|d_i\rangle \nonumber \\
&&= \left\vert\int^{\infty}_{\infty}\!\sqrt{\frac{2}{\pi w^2}}\,e^{-\frac{(x-d_i)^2+y^2}{w^2}}\varphi_{n0}(x,y) dx dy \right\vert^2
\nonumber \\
&&=\frac{d^{2n}_i}{w^{2n} n!}\exp\left(-\frac{d^2_i}{w^2}\right).
\label{eq:PerfectDetProb}
\end{eqnarray}
these distributions for the first several modes are shown as solid curves in Fig.\ref{fig:TheoryP} (a). Similarly, using a POVM description of a real detector (\ref{eq:POVM}) one obtains:
\begin{equation}
P_{d_i,n}=\exp(-d^2_i)\sum_{k,p=0}^{M}\frac{d^{k+p}_i}{\sqrt{k!p!}}\theta^{(n)}_{kp}=\sum_{k,p=0}^{M}F_{i,kp}\Pi_{kp,n},
\label{eq:ProbDistrVsPOVM}
\end{equation}
where we have introduced tensors $F_{i,kp}$ and $\Pi_{kp,n}$ as
$$
F_{i,kp}=\exp(-d^2_i)\frac{d^{k+p}_i}{\sqrt{k!p!}} ;\quad
\Pi_{kp,n}=\theta^{(n)}_{kp}.
$$
Our task now is to reconstruct the tensor of unknown coefficients $\Pi\in\mathbb{C}^{M\times M \times N}$ from the experimentally measured probability matrix $P\in\mathbb{R}^{D\times N}$, where $N$ is the number of detection modes, for which the data were taken in the experiment, $M$ is the number of modes used in the POVM decomposition (\ref{eq:POVM}), determining the number of free parameters for the fit, and $D$ is the number of discrete displacements of the input Gaussian beam. Let us note, that similarly to the situation described in \cite{lundeen2008tomography, feito2009measuring}, we find that the simplified rank-2 tensor $\Pi_{k,n}=\theta^{(n)}_{kp}\delta_{kp}$ gives a good fit of the experimental data. In what follows we use this simplified "diagonal" POVM form. As shown in \cite{lundeen2008tomography, feito2009measuring} the reconstruction procedure reduces to solving a constrained optimization problem:
\begin{equation}
\min\| P-F\Pi\|, \quad \Pi_n\geq 0, \quad \sum_{n=0}^{N-1}\Pi_n=\mathds{1},
\end{equation}
where the matrix norm is defined as $\|A\|=\sqrt{\sum_{i,j}|A_{i,j}|^2}$. The constraints are imposed on $M\times M$-dimensional matrices $\Pi_n$.



\section{Experimental results}

\begin{figure}[h]
\center{\includegraphics[width=0.6\linewidth]{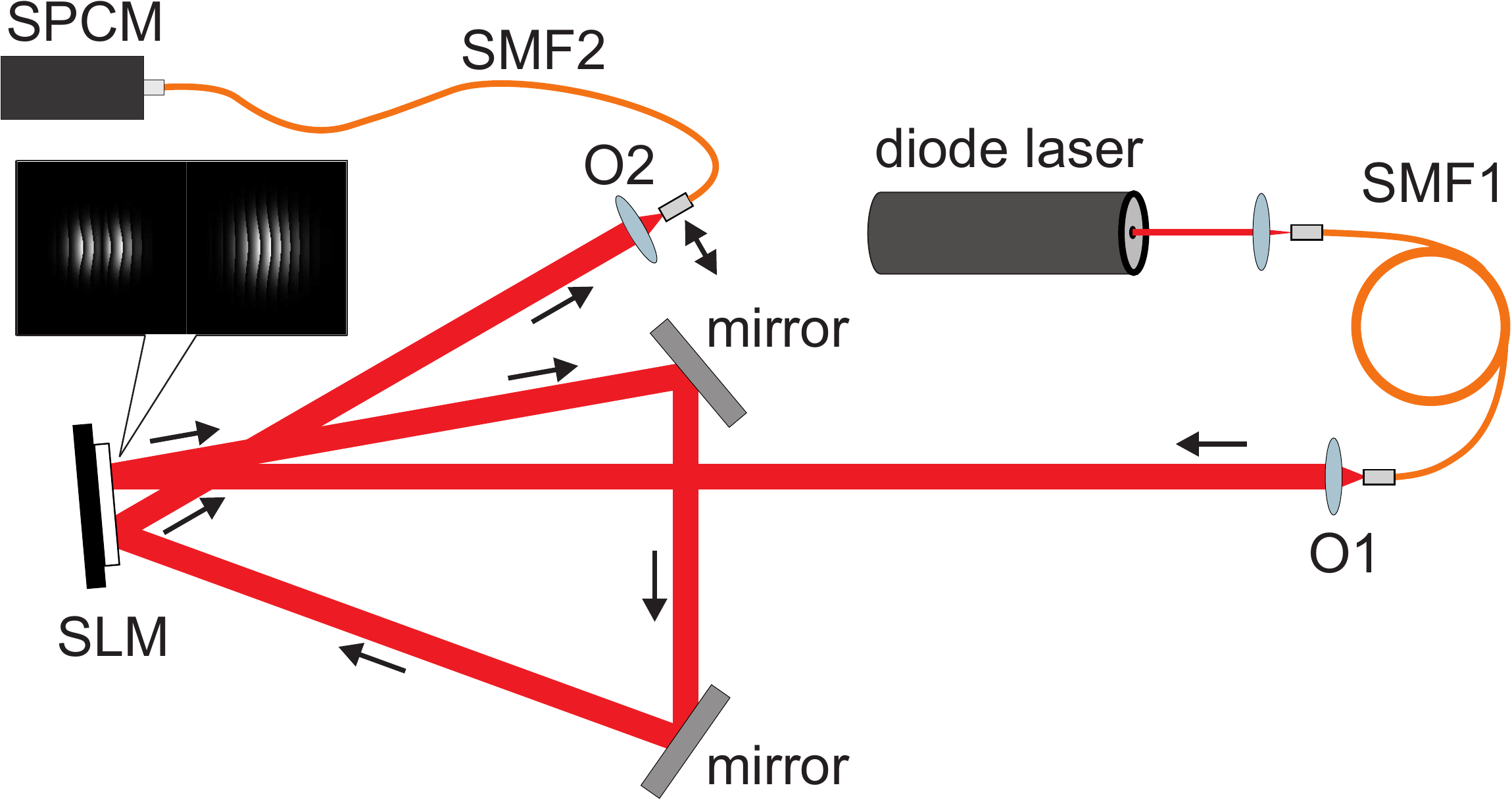}}
\caption{Experimental setup. A single phase-only SLM is used for both calibration beam preparation and as a part of the mode filter. The waist of the Gaussian beam is carefully controlled and mode-matched to the detection mode by the first phase hologram (right one on the inset). The mode-transforming hologram is displayed on the other part (left on the inset). The transformed beam is focused to a single mode fiber. }
\label{fig:ExpSetup}
\end{figure}

The scheme of the experimental setup is shown in Fig.\ref{fig:ExpSetup}. Attenuated radiation of a CW laser diode with 650 nm wavelength was mode-cleaned by a single mode fiber (SMF1) and collimated by a 8X microscope objective (O1). The waist of the beam was precisely controlled by a soft Gaussian aperture, realized as an amplitude-modulating hologram displayed on the right half of a phase-only SLM (Holoeye Pluto BB). The beam in the first diffraction order was reflected to the other half of the same SLM, which was used to display detection holograms. The first order of the detection part was mode-matched to a single-mode fiber (SMF2) with a 20x microscope objective (O2). SMF2 was placed in the back focal plane of the objective, realizing far-field detection. Finally, the intensity after SMF2 was measured with a single-photon counting module (SPCM).

For the sake of simplicity the tomographic procedure was described in the previous section for the case of near-field detection. Far field detection may be practically more advantageous, especially in the information transmission tasks, where collimated beams are required. However, since HG modes are eigenmodes of paraxial free propagation, the procedure is straightforwardly generalized to the case of far-field detection with the replacement of displaced Gaussian beams (\ref{eq:DispalcedGaussian}) by tilted Gaussian beams:
\begin{equation}
\varphi_{00}(x,y)=\sqrt{\frac{2}{\pi w^2}}\exp\left(-\frac{x^2+y^2}{w^2} \right)e^{i k_{i}x},
\end{equation}
where $k_{i}=2\pi\theta_i/\lambda$ is the transverse wave-vector component for a tilt angle $\theta_i$. Practically it is more convenient to shift the fiber tip in the focal plane of a focusing objective, than to tilt the probe beam. Obviously, the dependence of the detection probability for HG modes will still be expressed by (\ref{eq:PerfectDetProb}) with an appropriate scaling of waists.

\begin{figure}[h]
\begin{minipage}[h]{0.49\linewidth}
\center{\includegraphics[width=1\linewidth]{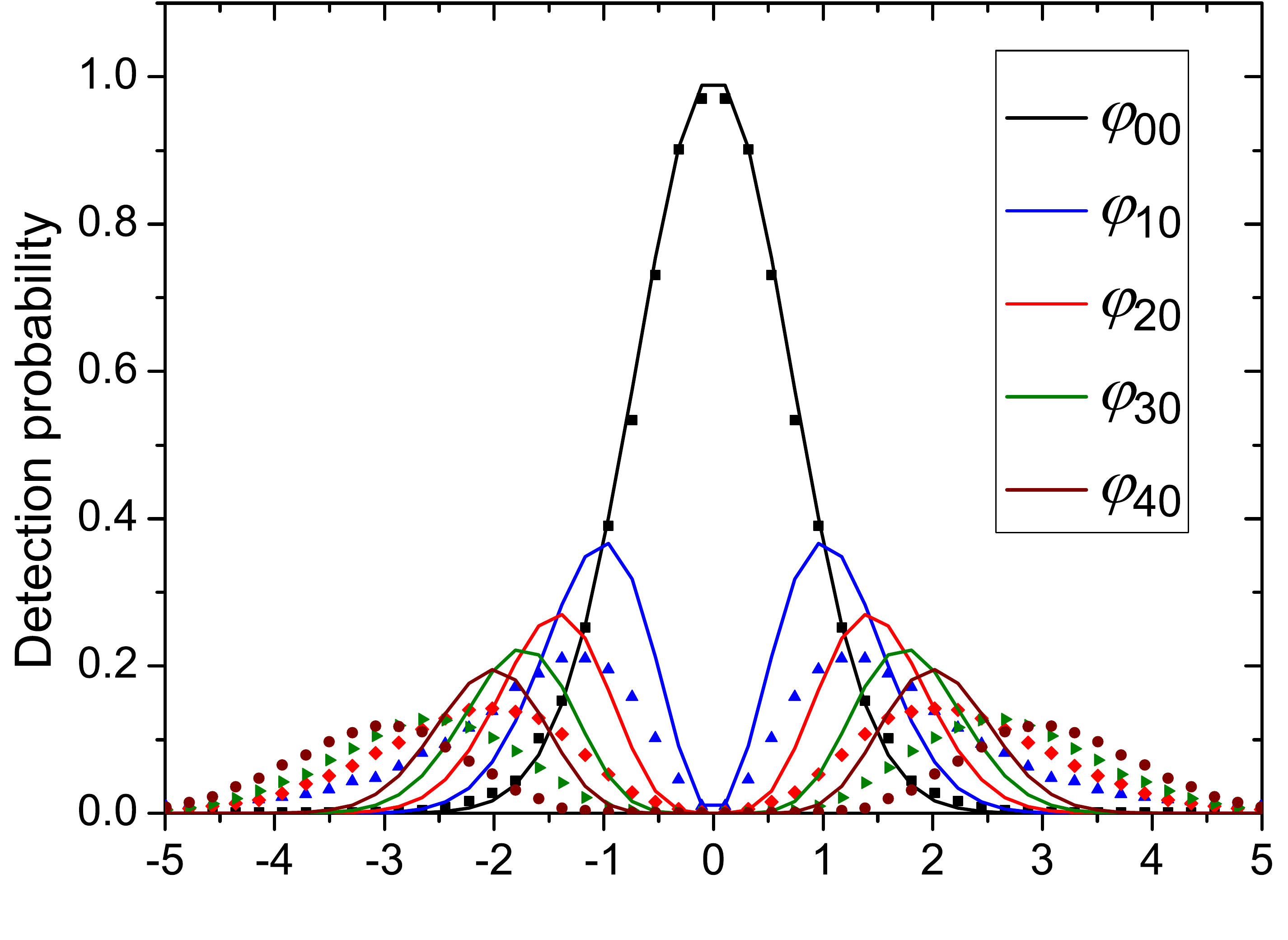} \\ a)}
\end{minipage}
\hfill
\begin{minipage}[h]{0.49\linewidth}
\center{\includegraphics[width=1\linewidth]{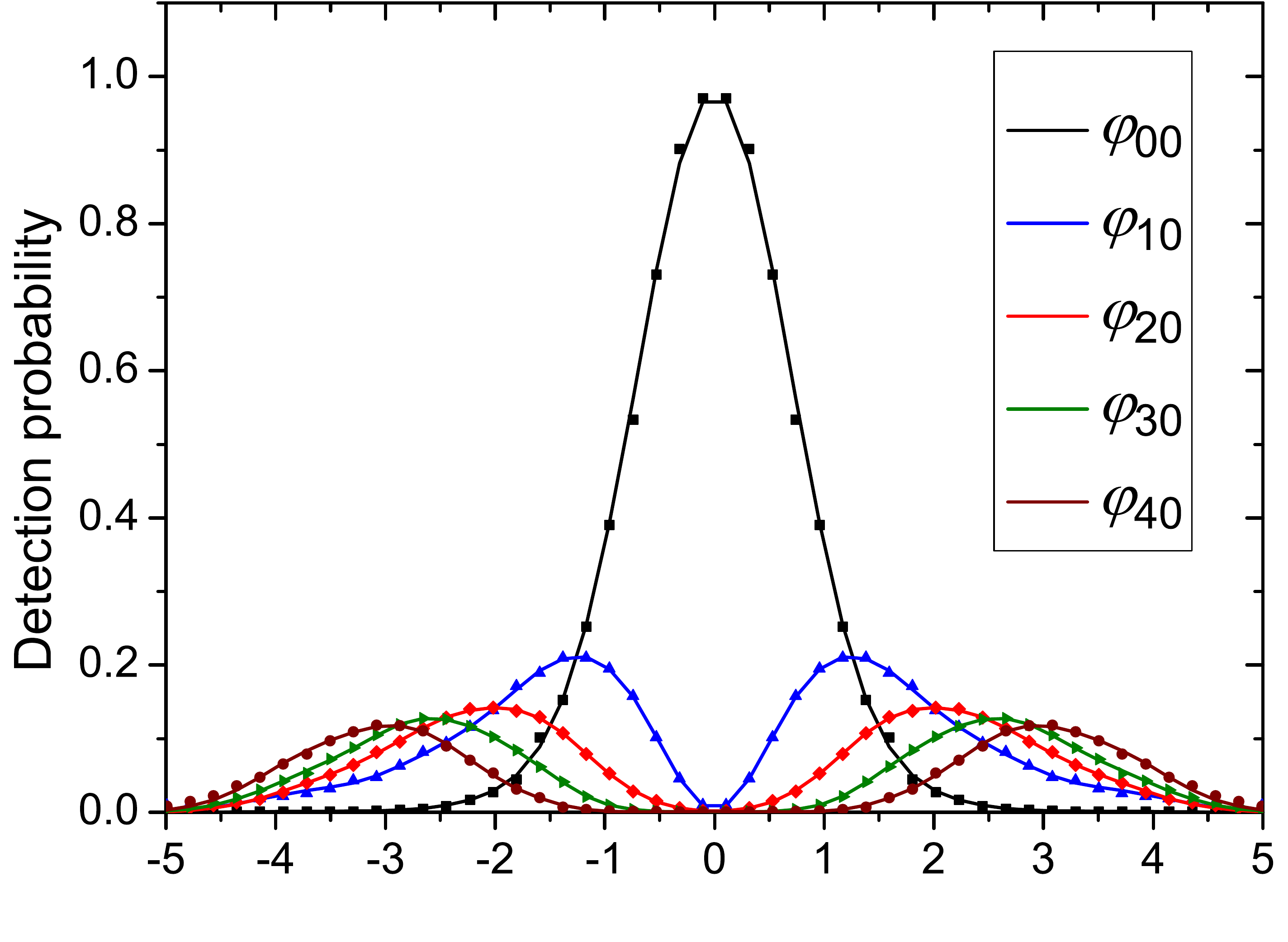} \\ b)}
\end{minipage}
\caption{Detection probability distributions for five lower-order phase-only holograms, corresponding to $n=0\ldots 4$ and $m=0$. Horizontal axis represents the displacement of the detecting fiber in the far field of the hologram $\delta_i$, corresponding to detection mode tilts $\theta_i=\delta_i/f$, where $f=8 mm$ is the focal length of the focusing objective. Displacement is given in dimensionless units $\delta_i/w$, where $w=(1.871 \pm 0.007) \mu$m is the fiber mode waist. Points are experimentally measured data for holograms with no amplitude modulation, solid lines are theoretical distributions for ideal HG modes (a) and probability distributions for reconstructed POVM of the real detector (b).}
\label{fig:TheoryP}
\end{figure}

\begin{figure}[h]
\center{\includegraphics[width=0.4\linewidth]{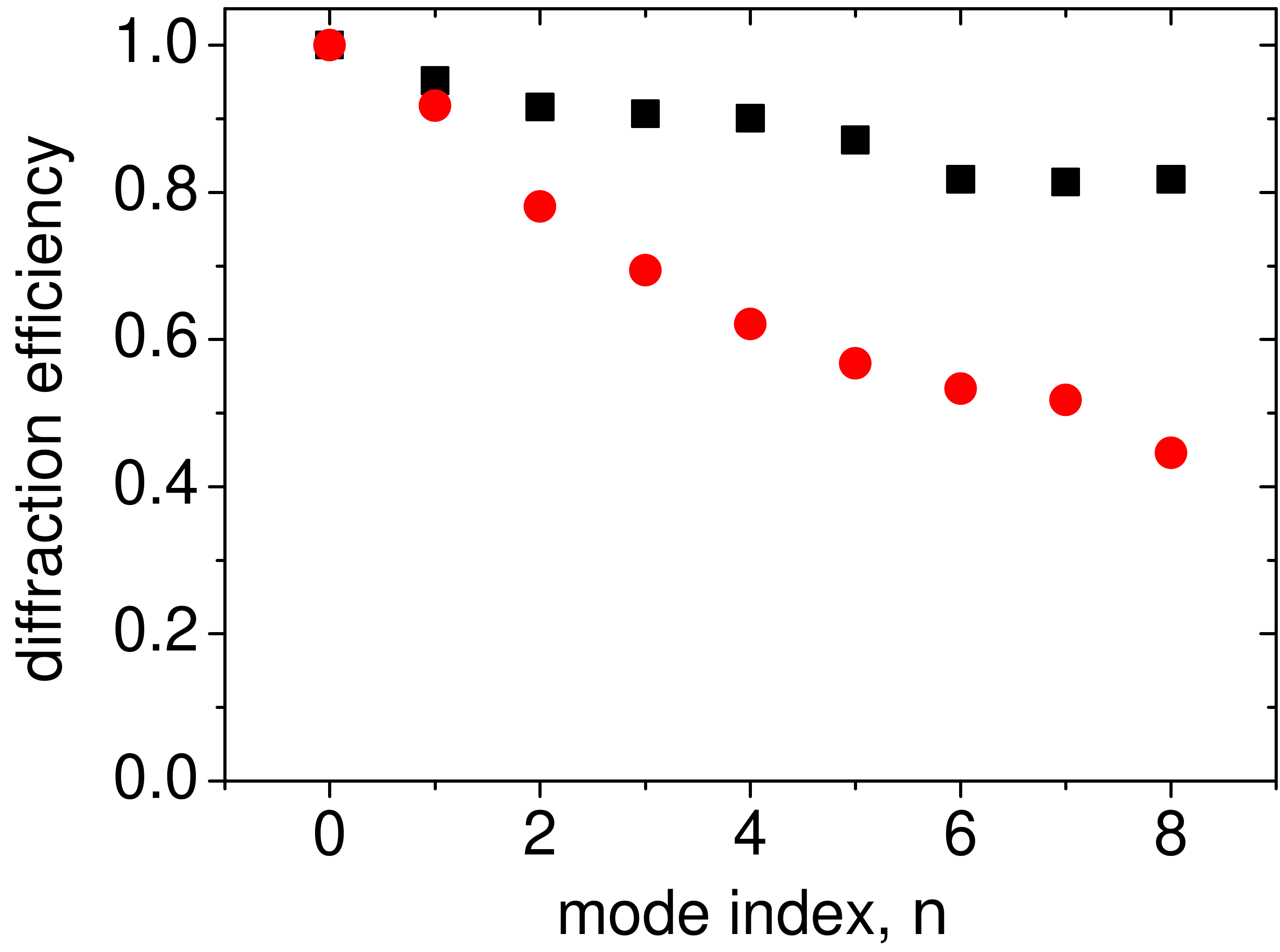}}
\caption{Diffraction efficiency measured as the ratio of integral intensity after the focusing objective to the intensity of the probe beam before the hologram. Black squares -- phase modulation; red circles -- phase and amplitude modulation. All data are normalized to the efficiency of the fundamental gaussian mode.}
\label{fig:Efficiency}
\end{figure}

\begin{figure}[h]
\begin{minipage}[h]{0.22\linewidth}
\center{\includegraphics[width=1\linewidth]{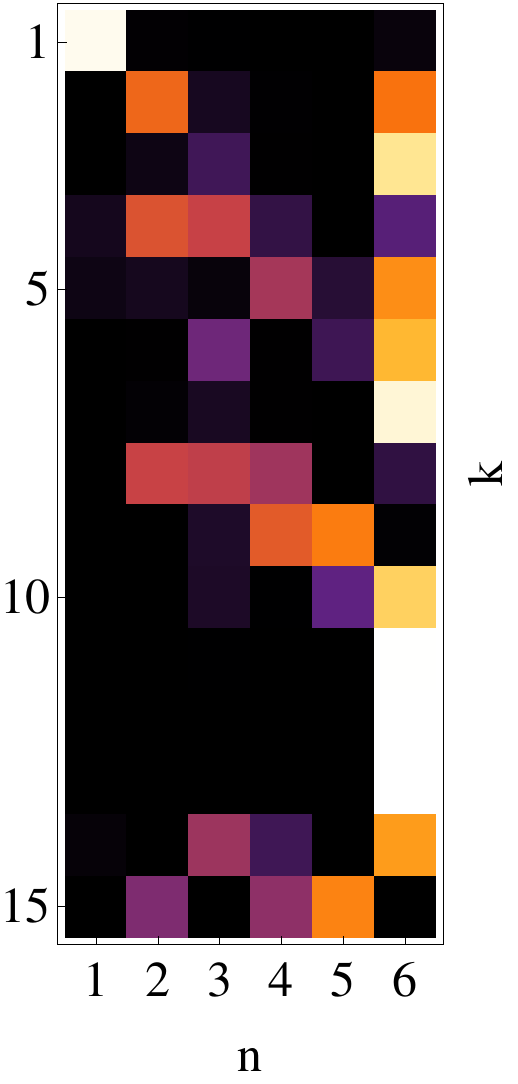}}
a) \\
\end{minipage}
\hfill
\begin{minipage}[h]{0.22\linewidth}
\center{\includegraphics[width=1\linewidth]{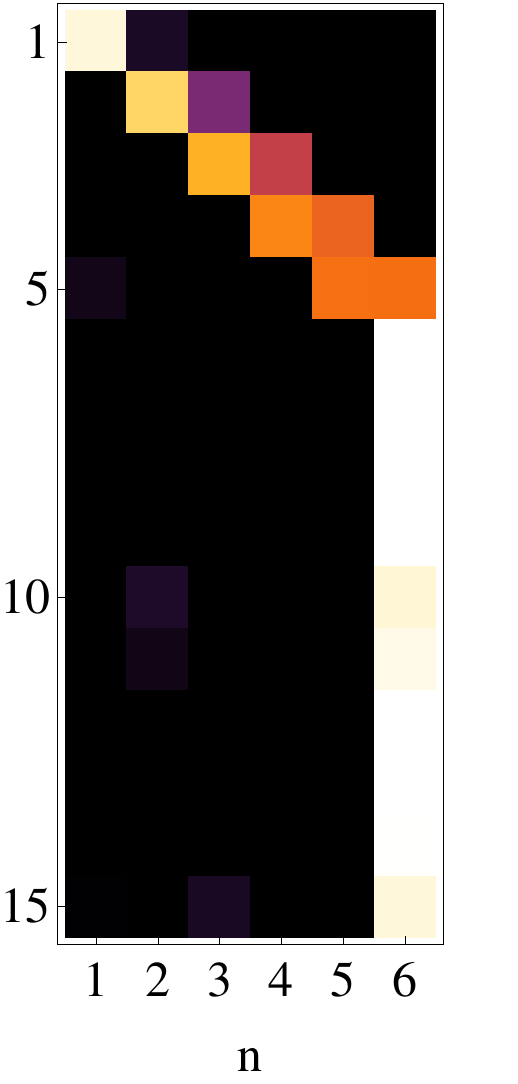}}
b) \\
\end{minipage}
\hfill
\begin{minipage}[h]{0.06\linewidth}
\center{\includegraphics[width=1\linewidth]{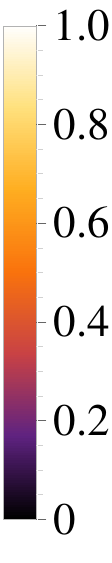}}
\end{minipage}
\hfill
\begin{minipage}[h]{0.22\linewidth}
\center{\includegraphics[width=1\linewidth]{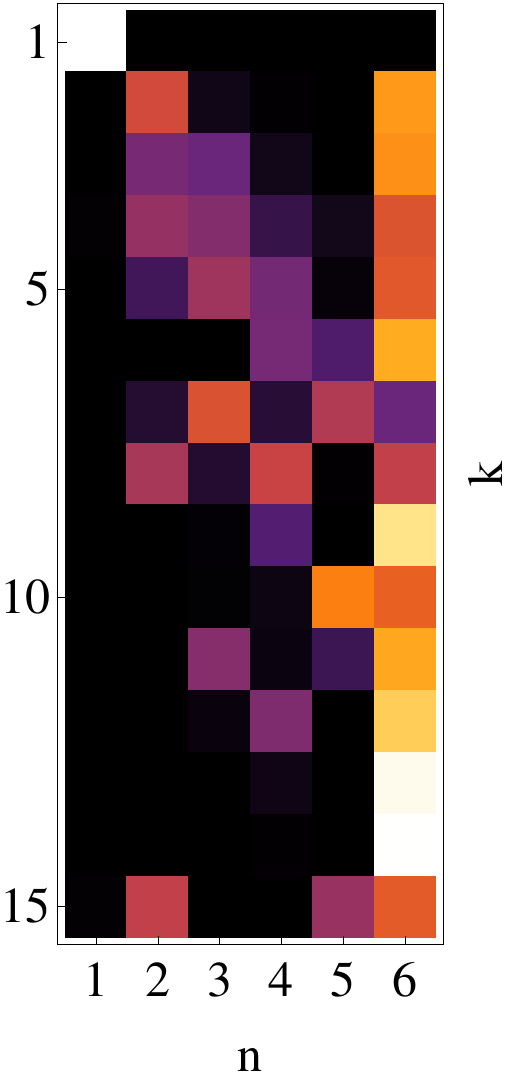}}
c) \\
\end{minipage}
\hfill
\begin{minipage}[h]{0.22\linewidth}
\center{\includegraphics[width=1\linewidth]{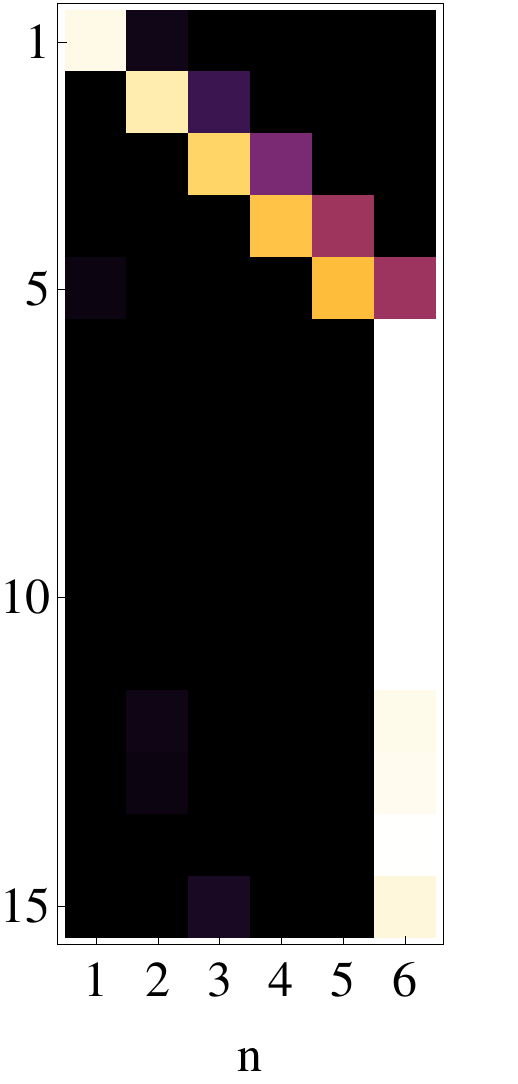}}
d) \\
\end{minipage}
\caption{Diagonal elements of the reconstructed POVM matrix in HG basis $\Pi_{n,kk}$: reconstructed from experimental data for holograms with no amplitude modulation (a), and with both phase and amplitude modulation (b); reconstructed from numerical simulations of the far-field diffraction pattern for holograms without (c) and with (d) amplitude modulation. }
\label{fig:POVMMatrices}
\end{figure}

To test the proposed tomographic procedure we have applied it to two variants of spatial mode filters: one with phase-only masks, and one with both phase and amplitude modulation given by the expressions (\ref{eq:ExactMask}). We expect the reconstruction to show superior performance of the second type of holograms, since phase-only masks are known to produce far-field intensity distributions different from those of ideal HG modes. These experimentally measured distributions for five lowest order modes are shown in Fig.~\ref{fig:TheoryP}. To take into account the non-unity diffraction efficiency, which is also hologram-dependent as shown in Fig.~\ref{fig:Efficiency}, we normalize the experimental distributions such that the integral intensity is unity for every mode. Comparing the experimental data with theoretical distributions expected for HG modes (solid lines in Fig.~\ref{fig:TheoryP}(a)), one can clearly see the slower decay for large displacements. The POVM elements obtained as a result of the reconstruction are shown in Fig.~\ref{fig:POVMMatrices}(a) (since the complete tensor is rank-3 we only show a two-dimensional part $\theta^{(n)}_{kk}$). The discrepancy between the detection modes and ideal HG modes is reflected in the presence of significantly large non-diagonal components. Using the reconstructed values for $\theta^{(n)}_{kp}$ we have calculated the detection probability distributions expected for the real detector, they are shown as solid lines in Fig.~\ref{fig:TheoryP}(b). One can clearly see the much better agreement with the experimental data. Quantitatively, the square of Pearson's correlation coefficient for the fit is $R_{rec}^2=0.9992$, while for the ideal theoretical distributions it is only $R_{th}^2=0.8632$. Same procedure for the holograms with amplitude modulation results in $R_{rec}^2=0.9848$ and $R_{th}^2=0.9647$, showing much better, although still non-ideal, performance of this type of holograms. It is also obvious in the Fig.~\ref{fig:POVMMatrices}(b), where the reconstructed POVM matrix elements are shown.

Having the reconstructed POVM elements for the detectors at hand we can quantify the detection performance by comparison with the diagonal tensor $\tilde \theta^{(n)}_{k}=\delta_{nk}$ for the ideal projectors in HG basis. A good measure of "closeness" for the probability distributions is given by \emph{similarity}, which in our case is expressed as:
\begin{equation}
    S=\frac{\left(\sum\limits_{n,k}{\sqrt{\theta^{(n)}_{k}\tilde\theta^{(n)}_{k}}}\right)^2} {{\sum\limits_{n,k}\theta^{(n)}_{k}}{\sum\limits_{n,k}\tilde\theta^{(n)}_{k}}}.
\end{equation}
The similarities of POVM's for detectors with phase-only and phase and amplitude holograms are $S_{ph}= 0.1937$ and $S_{amp}=0.7331$, respectively.

To be sure, that the observed features of POVM's for spatial filters are not artifacts of our setup, we have performed numerical simulations of these type of detectors. The far field distributions were calculated by fast fourier transform and numerically convolved with displaced Gaussian functions. The dependencies of the detection probabilities on the displacement, obtained numerically were subjected to the same reconstruction procedure as the experimental ones. The results are shown in Fig.~\ref{fig:POVMMatrices}(c) and Fig.~\ref{fig:POVMMatrices}(d) for phase-only and amplitude detection, respectively. The characteristic features of POVM's are reproduced in numerical simulations. The similarities between experimental and simulated POVM's are 0.8639 and 0.979 for phase-only and amplitude and phase masks respectively.

\section{Conclusions}
We have introduced a method for evaluation of the performance of holographic mode filters, which are widely used in quantum optical experiments with spatial entanglement, spatial encoding in classical and quantum optical communication and various other tasks. The advantage of the method is that it does not rely on any explicit calculations of the field transformation and regards the filter as a black-box which has to be described in the basis of HG modes. Thus it automatically reveals any systematic errors, whatever their reason is -- inappropriate hologram design, poor mode-matching, optical aberrations or anything else.

Of course, having a POVM description of the detector in HG basis, one can calculate its response on any other input state (for example, LG or other popular optical modes) algebraically, so the method is general and not limited to HG modes filters only. However, if one is interested in a particular set of modes, the set of calibration states may be optimized for this particular detector, as it was optimized here for filters of HG modes. For example, for LG modes it seems reasonable to use Gaussian beams with variable waists as calibration states. Results of \cite{qassim2014limitations} have already revealed imperfections of LG modes detection of the same origin as discussed here. Careful application of tomographic techniques to these data may provide further insight.

We have also provided another confirmation for the necessity to use amplitude modulation to design high-quality holographic mode filters. So the tradeoff between detection quality and detection efficiency seems to be inevitable for this type of filters. Other ways of detection are to be developed for the experiments where efficiency is crucial. Some alternatives exist, for example for separation of components with different values of orbital angular momentum \cite{berkhout2010efficient}, however their generalization to sorting of arbitrary orthogonal modes is highly non-trivial and requires additional research. We hope that methods developed here will find applications in these studies, which are of major importance for numerous applications of transverse modes of light.

\end{document}